\documentclass[journal]{IEEEtran}
\ifCLASSINFOpdf
\else
   \usepackage[dvips]{graphicx}
\fi
\usepackage{url}
\usepackage{amsmath}
\usepackage{mathtools}
\usepackage{multicol, multirow}
\usepackage{graphicx}
\usepackage{booktabs}
\usepackage{amssymb}
\usepackage{cite}
\usepackage{algorithm}% http://ctan.org/pkg/algorithm
\usepackage{algpseudocode}% http://ctan.org/pkg/algorithmicx
\usepackage{url}

\begin{document}

\title{Efficient Parallel Audio Generation using Group Masked Language Modeling}

\author{Myeonghun Jeong~\IEEEmembership{{Student Member,~IEEE,}} Minchan Kim~\IEEEmembership{{Student Member,~IEEE,}} Joun Yeop Lee, and Nam Soo Kim~\IEEEmembership{Senior Member,~IEEE}

\thanks{This work was supported by Samsung Research, Samsung Electronics Co.,Ltd.}
\thanks{Myeonghun Jeong, Minchan Kim, and Nam Soo Kim are with the Department of Electrical and Computer Engineering and with the Institute of New Media and Communications, Seoul National University, Seoul 08826, South Korea (e-mail: mhjeong@hi.snu.ac.kr; mckim@hi.snu.ac.kr; nkim@snu.ac.kr)}
\thanks{Joun Yeop Lee is with Samsung Research, Seoul, 06765, Republic of Korea (e-mail: jounyeop.lee@samsung.com)
}}

\markboth{Journal of \LaTeX\ Class Files, Vol. 14, No. 8, August 2015}
{Shell \MakeLowercase{\textit{et al.}}: Bare Demo of IEEEtran.cls for IEEE Journals}
\maketitle

\begin{abstract}
We present a fast and high-quality codec language model for parallel audio generation. While SoundStorm, a state-of-the-art parallel audio generation model, accelerates inference speed compared to autoregressive models, it still suffers from slow inference due to iterative sampling. To resolve this problem, we propose Group-Masked Language Modeling~(G-MLM) and Group Iterative Parallel Decoding~(G-IPD) for efficient parallel audio generation. Both the training and sampling schemes enable the model to synthesize high-quality audio with a small number of iterations by effectively modeling the group-wise conditional dependencies. In addition, our model employs a cross-attention-based architecture to capture the speaker style of the prompt voice and improves computational efficiency. Experimental results demonstrate that our proposed model outperforms the baselines in prompt-based audio generation.

% Scaling speech generation models to a large scale requires capturing the diversity of human speech, including speaker identities, prosodies, and acoustic details. Current large speech generation model systems typically quantize speech into discrete tokens to reduce the modeling complexity.

% and use autoregressive language models to generate these tokens, which can lead to unstable prosody and slow inference speed. 

% The generated acoustic tokens are then decoded by a HiFi-Codec to reconstruct the waveform. SoundOASIS is based on SoundStorm, but it improves the performance of SoundStorm to enable faster and higher-quality audio generation. The main differences between SoundStorm and SoundOASIS are as follows: 1) It uses a HiFi-Codec that is suitable for speech generation to obtain high-quality speech. 2) It improves the sampling speed without sacrificing the quality of the synthesized speech by using an order-agnostic sampling technique when combining the HiFi-Codecs. 3) It improves the inference speed while maintaining the conditioning power by using a speech prompting method with cross-attention. Experimental results show that the proposed model outperforms other audio generation baseline models and achieves superior performance on downstream tasks such as audio generation and zero-shot voice conversion.
\end{abstract}


\IEEEpeerreviewmaketitle

\section{Introduction}
% 1031  첫 문단만 수정하면 될듯

% \IEEEPARstart{N}{eural} audio generation is a promising technology with the potential to revolutionize a wide range of industries. For example, \cite{oord2016wavenet, kong2020hifi} successfully reconstruct audio from the acoustic feature (i.e., mel-spectrogram), and \cite{kim2021conditional} synthesize high-quality speech from the text. \cite{dhariwal2020jukebox} generates music exploiting musical notes and lyrics. Such approaches use copious information about the underlying audio content to adapt the model to versatile downstream tasks (i.e., speech and music generation) 

\IEEEPARstart{R}{ecent}
development of neural audio codecs~\cite{defossez2022highfi, zeghidour2021soundstream} has brought significant attention to large language models~(LLM) as a promising avenue for audio generation. The transformer-based LLMs in Natural Language Processing~(NLP) area have demonstrated their outstanding performance by capturing the long-term context and remarkable zero-shot capability through in-context learning~\cite{brown2020language, touvron2023llama}. Inspired by this line of research, casting the audio generation in the continuous domain~\cite{oord2016wavenet,kong2020hifi,kong2020diffwave} to the discrete domain, by taking advantage of a powerful LLM, has unlocked rapid progress in versatile applications.
% the discrete audio representation from neural audio codecs~\cite{yang2023hifi,defossez2022highfi, zeghidour2021soundstream} has led to a paradigm shift, from the audio generation task to the transformer-based~\cite{vaswani2017attention} language modeling task.
% capturing the long-term context. Especially, the prefix language model utilizes prompt information, which provides tasks or speaker information in the audio generation tasks~[...].
% prompt 얘기
Notably,~\cite{borsos2023audiolm} introduced an autoregressive transformer to model the discrete acoustic tokens, exploring its application in audio continuation tasks by using the audio prefix as a prompt. Furthermore,~\cite{wang2023neural} and~\cite{kharitonov2023speak} successfully employed codec language models in zero-shot speech synthesis, using only a few seconds of an unseen prompt voice. Despite these advancements, the length of an acoustic token sequence generated from neural audio codecs is typically longer than that of natural language tokens due to its frame rate. This poses challenges for developing transformer-based discrete audio generation models that have quadratic runtime complexity.

To address this issue, prior research~\cite{borsos2023soundstorm, garcia2023vampnet, copet2023simple, shen2023naturalspeech, lan2023stack} proposed various methods to enhance computational efficiency. For instance, \cite{copet2023simple} and \cite{lan2023stack} suggested novel codebook patterns to reduce iterations in autoregressive modeling, while \cite{shen2023naturalspeech} introduced a non-autoregressive diffusion model~\cite{ho2020denoising} for modeling the continuous acoustic token embedding. SoundStorm~\cite{borsos2023soundstorm}, the primary focus of this work, introduced a confidence-based parallel decoding technique for modeling the discrete acoustic token sequence. Leveraging the characteristics of residual vector quantization~(RVQ)-based codebooks~\cite{zeghidour2021soundstream}, the confidence-based parallel decoding technique significantly reduced the complexity of non-autoregressive models, generating acoustic tokens iteratively with fewer sampling passes. Although these approaches have somewhat improved inference speed, they still show slow generation due to their iterative nature.

%To mitigate this problem,~\cite{borsos2023soundstorm, garcia2023vampnet, copet2023simple, shen2023naturalspeech} have introduced a non-autoregressive structure to expedite inference speed and enhance audio quality. 

%The SoundStorm~\cite{borsos2023soundstorm}, primary focus of this work, introduces the semantic to acoustic token translation model that relies on a bidirectional attention mechanism and confidence-based parallel decoding. The bidirectional attention effectively models long-term dependencies in prompt speech, leading to improved speaker consistency in generating long audio samples. The confidence-based parallel decoding reduces the modeling complexity of non-autoregressive models, enabling SoundStorm to generate samples with just 27 sampling passes. While reducing the number of iterations in parallel decoding can help lower computational costs compared to autoregressive models, it often comes at the expense of speech quality. This poses a dilemma, as the original SoundStorm frequently trades the sample quality to expedite sampling speed.

\begin{figure} 
\centering
  \includegraphics[width=0.75\columnwidth]{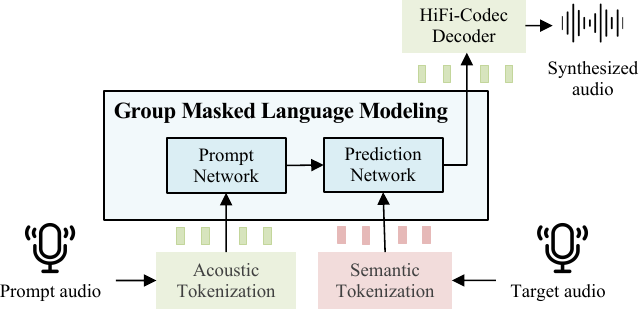}
  \centering
  \caption{Overview of our proposed model}
  \label{fig_overview}
  \vspace{-0.2cm}
\end{figure}

Motivated by this problem, we propose a fast, high-quality codec language model for parallel audio generation. As illustrated in Fig.~\ref{fig_overview}, our approach focuses on semantic-to-acoustic token generation given prompt acoustic tokens. We employ HiFi-Codec~\cite{yang2023hifi} for acoustic tokenization and Wav2Vec 2.0~\cite{baevski2020wav2vec} for semantic tokenization. HiFi-Codec provides Group-RVQ (G-RVQ)-based acoustic tokens, facilitating high-quality audio tokenization with more concise codebooks. Based on these G-RVQ acoustic tokens, we propose an efficient training algorithm, Group-Masked Language Modeling~(G-MLM), which employs group-wise conditional dependency. Furthermore, we propose Group-Iterative Parallel Decoding (G-IPD), mirroring this training procedure, and verify that G-IPD enables our model to generate acoustic tokens with fewer iterations without compromising audio quality. Additionally, we propose a cross-attention-based prompting method, a computationally efficient structure for reflecting the speaker identity of the prompt voice. 

% We demonstrate that our proposed model outperforms other baselines in zero-shot voice conversion task.

% The main contributions of our work are as follows:
% \begin{itemize}
% \item We propose the 
% \item We propose the order-agnostic sampling technique to expedite inference speed without sacrificing the audio quality.
% \item We firstly enjoy the cross attention scheme to condition the prompt speaker identity in predicting 
% \item We demonstrate that our proposed model can generate high fidelity audio and faster than original SoundStorm model.
% \end{itemize}

% In the case of unconditional modeling, the model handles acoustic token continuation tasks. This transformer-based audio generation model unlocked rapid development in speech continuation, text-to-speech, and music generation.

% 문단
% 1. audio generation 소개, application 소개
% 2. conditional audio generation, unconditional audio generation 소개 및 모델 몇가지 소개
% 3. acoustic token modeling을 할때 autoregressive 한 방식의 문제점 및 모델 소개
% 4. soundstorm의 소개 및 문제점
% 5. proposed model의 소개

%%%%%

% 포함 내용 : 
% 1. RVQ, neural audio coder
% 2. language model (prompt) 방식의 TTS 모델
% 2-1. conventional TTS (audio generation) 모델 (?)
% 3. audio generation model (AudioLM, Soundstorm 등) -> downstream task로 zero-shot TTS / voice conversion 등이 가능함을 설명

% \input{compare_tab}
\section{Backgrounds}

\subsection{Group Residual Vector Quantization (G-RVQ)}
Residual Vector Quantization~(RVQ), employed in SoundStream~\cite{zeghidour2021soundstream} and Encodec~\cite{defossez2022highfi}, encodes multiple streams of discrete tokens from audio, within the framework of VQ-VAE~\cite{van2017neural}. RVQ compresses each audio frame through cascaded quantizers, with each quantizer contributing residually to the encoding process, generating multi-level sequences of codewords. In this configuration, the initial level codebook retains the most fundamental audio information, and the number of quantization levels $N_q$ controls the trade-off between computational cost and coding efficiency.

More recently, HiFi-Codec~\cite{yang2023hifi} introduced a Group Residual Vector Quantization (G-RVQ) scheme, demonstrating superior performance at lower bit rates. G-RVQ divides the latent features extracted from the encoder into $G$ groups and applies RVQ to each group with $N_q$ levels. For example, with a target bitrate of $R=2000$ bps and 50 output frames per second, resulting in $r=2000/50 = 40$ bits allocated to each frame, and for $N_q=2$ and $G=2$, the total rate budget is evenly distributed among each Vector Quantization~(VQ) layer, i.e., $r_i = r/(N_q*G) = \log_2 N$. Consequently, the codebook size becomes $N=2^{r_i}=2^{40/4}=1024$. As G-RVQ utilizes multiple initial levels of RVQ codebooks, it demonstrates a higher compression rate compared to RVQ.

\begin{figure} 
\centering
  \includegraphics[width=0.80\columnwidth]{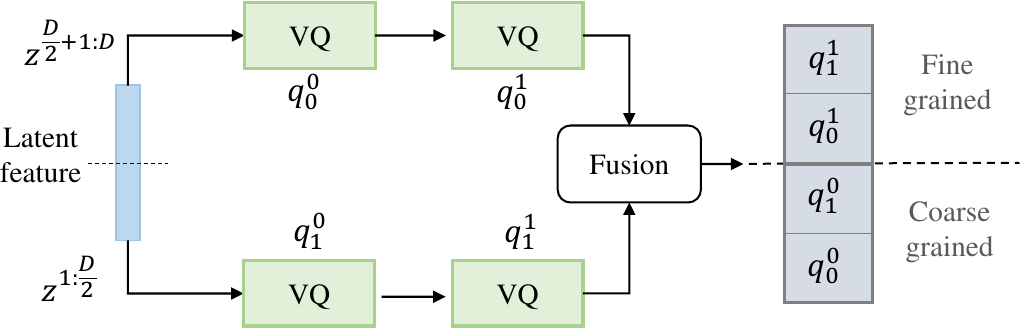}
  \centering
  \caption{Bi-group, bi-depth G-RVQ for acoustic tokenization}
  \label{fig_grvq}
  \vspace{-0.2cm}
\end{figure}

\subsection{SoundStorm}
SoundStorm is a non-autoregressive model designed for translating semantic tokens into acoustic tokens. Semantic tokens are derived from W2V-BERT~\cite{chung2021w2v} to encode coherent semantic information, while RVQ-based acoustic tokens are extracted from the SoundStream~\cite{zeghidour2021soundstream} audio codec for encoding acoustic information. SoundStorm~\cite{borsos2023soundstorm}, comprising conformer blocks~\cite{gulati20_interspeech}, is trained to predict masked acoustic tokens given the semantic tokens. The bidirectional conformer structure allows for acoustic token generation in an arbitrary order, ensuring prompt speaker and acoustic consistency. In order to improve the inference speed, SoundStorm applies the iterative sampling scheme of MaskGIT~\cite{chang2022maskgit} for parallel audio generation. At each sampling iteration, the top-$k$ predicted tokens with the highest confidence scores are kept fixed, while the rest are predicted again. The number of predicted tokens in each round is gradually increased, ensuring the conditional dependency between acoustic tokens and this process proceeds RVQ level-wise in a coarse-to-fine order. Although SoundStorm improves the inference speed compared to autoregressive models, it often compromises the speech quality when reducing the decoding iterations.

% SoundStorm sets the parameter $N_q$ to 12 and utilizes iterative parallel decoding only in the first quantization level of codebooks. The remaining codebooks are predicted in a single pass per RVQ level, resulting in a total of 27 forward passes.

% \begin{figure*} 
% \centering
%   \includegraphics[width=1.75\columnwidth]{model6_crop.pdf}
%   % \includegraphics[width=0.95\columnwidth]{model6_crop.pdf}
%   \centering
%   \caption{Overall training and sampling procedure are depicted on the left side. Model architecture is depicted on the right side. $s$ denotes the iteration steps.}
%   % \vspace{-0.5}
%   \label{fig_train}
% \label{training}
% \vspace{-0.2cm}
% \end{figure*}

\begin{figure} 
\centering
  \includegraphics[width=0.78\columnwidth]{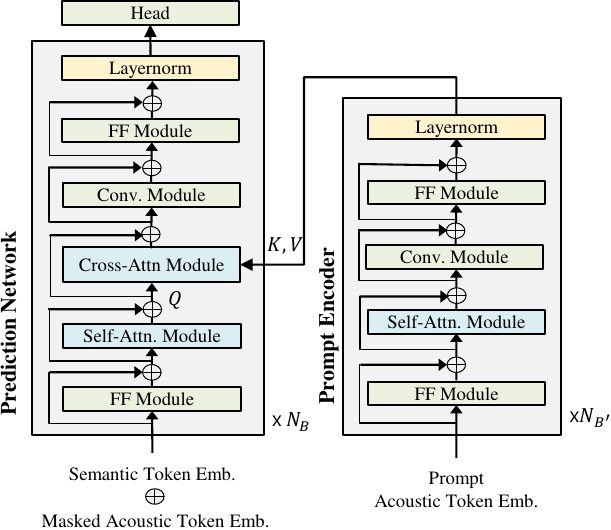}
  \centering
  \caption{Overall model architecture}
  % \vspace{-0.5}
  \label{fig_train}
\label{training}
\vspace{-0.2cm}
\end{figure}

\section{Proposed method}
\subsection{Tokenization}
We perform $k$-means clustering for semantic tokenization over the 15th hidden representation of Wav2Vec~2.0~\cite{baevski2020wav2vec}. Previous research~\cite {baevski2021unsupervised, kim22c_interspeech} showed that these discrete tokens effectively capture semantic information, even substituting phonetic sequences. For acoustic tokens, we leverage the bi-group bi-depth G-RVQ of HiFi-Codec as illustrated in Fig.~\ref{fig_grvq}. Let $\mathbf{x}$ represent a waveform, and $\mathbf{z}\in \mathbb{R}^D$ be a latent feature. The acoustic tokenization process is performed as follows:
\begin{equation}
\label{Acoustic token extraction}
\begin{split}
 \mathbf{z}^{1:D} &= Enc(\mathbf{x})  \\
 [\mathbf{q}^0_0, \mathbf{q}^0_1, \mathbf{q}^1_0, \mathbf{q}^1_1] &= GRVQ(\mathbf{z})\\
 \mathbf{q}_c, \mathbf{q}_f &= [\mathbf{q}^0_0, \mathbf{q}^0_1], [\mathbf{q}^1_0, \mathbf{q}^1_1],
\end{split}
\end{equation}
where $\mathbf{q}^j_g \in \{1, 2,..., C\}^{T}$ represents the acoustic token sequence for $j$-th quantizer level and $g$-th group. The maximum length and codebook size are denoted as $T$ and $C$, respectively. We denote $\mathbf{q}_c$ as the coarse-grained acoustic tokens and $\mathbf{q}_f$ as the fine-grained acoustic tokens. Utilizing G-RVQ rather than RVQ can encode more abundant acoustic information, and its concise RVQ depth brings computational efficiency.

\begin{figure*} 
\centering
  \includegraphics[width=2.0\columnwidth]{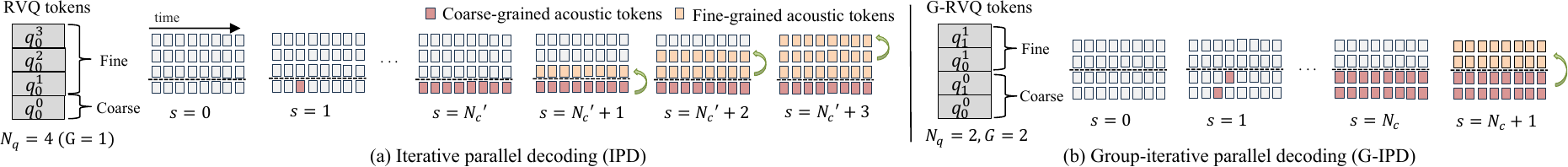}
  \centering
  \caption{Comparison of iterative inference process: (a) SoundStorm's IPD, and (b) proposed method's G-IPD technique. $s$ denotes the iteration steps.}
  % \vspace{-0.5}
  \label{fig_gipd}
\vspace{-0.2cm}
\end{figure*}

\subsection{Model Architecture}
As shown in Fig.~\ref{training} (a), our model builds upon the architecture of SoundStorm~\cite{borsos2023soundstorm}, employing a conformer~\cite{gulati20_interspeech} module and a masked language modeling approach~\cite{devlin-etal-2019-bert}. For the input of the prediction network, we aggregate the embeddings of the semantic tokens and the corresponding frames of partially masked acoustic tokens. Then, our model predicts acoustic tokens given the prompt acoustic token embedding as a conditioning signal. The output embeddings from the prediction network are processed by separate heads for each RVQ level.

To capture the speaker information of the prompt voice, we employ a multi-head cross-attention module~\cite{vaswani2017attention} in the prediction network, as shown in Fig.~\ref{fig_train} (a). 
% The query is derived from the hidden sequence of the self-attention layer, while the key and value are derived from the hidden sequence of the prompt encoder.
Let $\mathbf{e}$ denote the output of the prompt encoder and $\mathbf{h}$ be the output of the self-attention in the prediction network. The key, $\mathbf{K}$ and value, $\mathbf{V}$ are derived from $\mathbf{e}$, and the query, $\mathbf{Q}$ is obtained from $\mathbf{h}$. 
The multi-head cross-attention module operates as follows:
\begin{equation}
\label{Cross attention}
\begin{split}
 \mathbf{Q}_i = \mathbf{hW}_q^i&,\;  \mathbf{K}_i = \mathbf{eW}_k^i, \; \mathbf{V}_i = \mathbf{eW_v^i}   \\
 \mathbf{head}_i &= Softmax(\frac{\mathbf{Q}_i\mathbf{K}_i}{\sqrt{d}})\mathbf{V}_i\\
 \mathbf{c} &= [\mathbf{head}_1, ..., \mathbf{head}_{N_h}],
\end{split}
\end{equation}
where $\mathbf{W}^i_q$, $\mathbf{W}^i_k$, and $\mathbf{W}^i_v$ denote the linear projections for the key, value, and query, respectively. In~(\ref{Cross attention}), $\mathbf{c}$ represents the context vector summarizing the prompt voice. 
By leveraging the cross-attention mechanism~\cite{vaswani2017attention}, there are distinct advantages compared to SoundStorm.
% Firstly, the attention matrix of the cross-attention operates unidirectionally from the prompt to the target sequence, which provides robust conditioning. As a result, our prediction network retains the speaker and acoustic consistency of the prompt embedding, reducing redundant modeling complexity compared to self-attention. 
Firstly, unlike SoundStorm, which requires semantic tokenization of the prompt sequence, our model seamlessly captures prompt information using only acoustic tokens. This simplifies the inference process by eliminating the necessity for semantic tokenization of the prompt sequence. Secondly, our cross-attention mechanism strategically caches the key and value, avoiding the need for repetitive computation of the prompt part throughout the iterative sampling process. As a result, the prompt part needs to be calculated only once during inference while maintaining the prompt speaker information.

% (1) In contrast to SoundStorm, which mandates semantic tokenization of the prompt sequence, our model efficiently incorporates prompt information using only acoustic tokens. This simplifies the inference process by eliminating the necessity for semantic tokenization of the prompt sequence. (2) Unlike the self-attention structure in SoundStorm, which requires continual computation of the prompt part during the iterative sampling process, our cross-attention mechanism caches the key and value during inference. As a result, the prompt part only needs to be calculated once throughout the entire iterative sampling process, reducing computational redundancy.

\subsection{Training and Inference}
\label{section: training procedure}
To harness the full potential of G-RVQ acoustic tokens, we propose the Group-Masked Language Modeling (G-MLM) approach for training the model. In this training scenario, we first sample the prompt delimiter time step $t \sim U[\epsilon, T-1]$ to separate prompt and target sequence, where $\epsilon$ means a starting frame index. The tokens before $t$ constitute the prompt acoustic tokens, while those after $t$ form the target sequence for generation. Our core idea lies in the masking strategy outlined in~\textbf{Algorithm~\ref{G-MLM}}. When training the coarse-grained acoustic tokens $\mathbf{q}_c=[\mathbf{q}^0_0, \mathbf{q}^0_1]$, we apply the cosine-scheduling mask~\cite{borsos2023soundstorm, chang2022maskgit} separately to $\mathbf{q}^0_0$ and $\mathbf{q}^0_1$ in temporal axis, while masking all of the fine-grained acoustic tokens. As the G-RVQ tokens are extracted from the same latent feature, we assume that $\mathbf{q}^0_0$ and $\mathbf{q}^0_1$ are highly entangled. Based on this assumption, our inter-group masking strategy reduces the modeling complexity by employing the group-wise conditional dependency. For training fine-grained acoustic tokens, we apply the cosine mask to the fine-grained acoustic tokens in the same manner. Our model exploits RVQ-depth-wise conditional dependency in this step by leaving coarse-grained acoustic tokens unmasked. Finally, our model is trained with the cross-entropy loss using the ground-truth acoustic tokens as the target, and the loss is calculated only for the masked tokens.

For inference, we propose the Group Iterative Parallel Decoding (G-IPD) technique, mirroring the G-MLM training scheme.
% Our sampling technique leverages the property that the codebooks of different groups with the same quantization level are entangled with each other and exhibit no inherent order. Therefore, our model can operate by expanding the search space by a factor of the number of groups at each iteration.
%Since our model is trained to predict coarse or fine-grained codebooks regardless of their order, OASIS models the conditional dependency within the coarse or fine-grained codebooks. Our model architecture which consists of parameter-sharing conformer but only different heads facilitate this efficient training. In inference stage, our model samples the tokens in a coarse-to-fine order. By doubling the search space, we expect that the performance degradation is small even if the number of iterations is reduced. 
%To support this inference scheme, we propose a masking scheme for training that mimics the inference procedure.
Fig.~\ref{fig_gipd} illustrates a comparison between G-IPD and SoundStorm's IPD~\cite{borsos2023soundstorm}. Both techniques initially predict coarse-grained acoustic tokens and subsequently fine-grained acoustic tokens. When predicting coarse-grained acoustic tokens, a confidence-based iterative sampling~\cite{chang2022maskgit, garcia2023vampnet, borsos2023soundstorm} scheme is employed. At each iteration, the predicted tokens with the highest confidence scores are fixed, while the rest are re-masked. The number of masked tokens for each round is gradually decreased, following cosine schedule~\cite{borsos2023soundstorm}. The key difference from the SoundStorm~\cite{borsos2023soundstorm}'s IPD is that our decoding scheme involves the acoustic token sequences from two distinct groups together in the search space for each iteration. Doubling the search space allows our model to exploit group-wise conditional dependency, resulting in fewer iterations without performance degradation. Furthermore, G-RVQ inherently encodes rich audio information even at lower bitrates than RVQ. This contributes to faster inference while keeping the audio quality. Once the coarse-grained acoustic tokens are generated, they are used as conditions for predicting fine-grained acoustic tokens in a single step.

\label{sec:guidelines}

% \begin{algorithm}
%   \caption{Group Masked Language Modeling (G-MLM)}
%   \label{G-MLM}
%   \textbf{Input} model $g_\theta$, semantic tokens $\mathbf{\rho}$, prompt acoustic tokens $\mathbf{\hat{q}}$ \\
%   \textbf{Output} The loss  $\mathcal{L}$
%   \begin{algorithmic}[1]
%   \State $l \sim Bernouli(0.5)$ \Comment{Sample quantization level}
%   \If{$l=0$}  \Comment{Training the coarse-grained codebooks}
%   \State $\textbf{q}_{c}^M=Cosine Mask(\textbf{q}_{c})$, 
%   \State $\textbf{q}_{f}^M=Entire Mask(\textbf{q}_{f})$
%   \Else   \Comment{Training the fine-grained codebooks}
%   \State $\textbf{q}_{f}^M=Cosine Mask(\textbf{q}_{f})$
%   \EndIf
%   \State $\textbf{q}^M= Concat(\textbf{q}_c^M, \textbf{q}_f^M)$
%     \State $ p_{\theta}(\mathbf{q}|\mathbf{q}^M, \mathbf{\hat{q}}, \mathbf{\rho}) = Softmax(g_\theta(\mathbf{q}^{M},\mathbf{\hat{q}},\mathbf{\rho}))$ 
%     \State $\mathbf{\tilde{y}} \sim p_{\theta}(\mathbf{q}|\mathbf{q}^M, \mathbf{\hat{q}}, \mathbf{\rho})$

% \State $\mathcal{L} = CrossEntropy(\mathbf{\tilde{y}}$,$\mathbf{q})$ \Comment{Calculated only for cosine-masked tokens}
% \State \textbf{return} $\mathcal{L}$
%   \end{algorithmic}
% \end{algorithm}

\begin{algorithm}
  \caption{Masking strategy for G-MLM}
  \label{G-MLM}
  \textbf{Input} coarse-grained acoustic tokens $\textbf{q}_{c}$, fine-grained acoustic tokens $\textbf{q}_{f}$ \\
  \textbf{Output} masked acoustic token $\mathbf{q}^M$
  \begin{algorithmic}[1]
  \State $l \sim Bernoulli(0.5)$ \Comment{Sample quantization level}
  \If{$l=0$}  \Comment{Training the coarse-grained acoustic tokens}
  \State $\textbf{q}_{c}^M=Cosine Mask(\textbf{q}_{c})$, 
  \State $\textbf{q}_{f}^M=Entire Mask(\textbf{q}_{f})$
  \Else   \Comment{Training the fine-grained acoustic tokens}
  \State $\textbf{q}_{f}^M=Cosine Mask(\textbf{q}_{f})$
  \EndIf
  \State $\textbf{q}^M= Concat(\textbf{q}_c^M, \textbf{q}_f^M)$

\State \textbf{return} $\mathbf{q}^M$
  \end{algorithmic}
\end{algorithm}
\vspace{-0.3cm}

% \begin{algorithm}
%   \caption{Group Iterative Parallel Decoding (G-IPD)}
%   \label{G-IPD}
%   \textbf{Input} model $g_\theta$, semantic token $\mathbf{t}_s$, prompt acoustic token $\mathbf{t}_a$, coarse-grained acoustic token $\mathbf{q}_c$  \\
%   \textbf{Output} predicted coarse-grained acoustic token $\mathbf{\tilde{q}}_{c}$
%   \begin{algorithmic}[1]
%       \For{$n \leftarrow \; 0 \; to \; N-1$}
%         \State $ \mathbf{q}_{c,n+1} = g_\theta(\mathbf{q}^{M}_{c,n},\mathbf{t}_s, \mathbf{t}_a)$ 
%         % \State $\epsilon \sim Gumbel(0,1)$ 
%         % \State $y \leftarrow y + \tau \cdot \epsilon$  \Comment{Add Gumbel noise to logits}
%         % \State $p_{\theta}(\mathbf{q}_{c,n+1}|\mathbf{q}_{c,n}^M, \mathbf{t}_a, \mathbf{t}_s) \leftarrow Softmax(y)$ 
%         % \State $\mathbf{q}_{c,n+1} \sim p_{\theta}(\mathbf{q}_{c,n+1}|\mathbf{q}_{c,n}^M, \mathbf{t}_a,\mathbf{t}_s)$
%         \State $\mathcal{C} \leftarrow p_{\theta}(\mathbf{q}_{c,n+1}|\mathbf{q}_c^M, \mathbf{t}_a,\mathbf{t}_s) $ 
%         \State Rank by confidence $\mathcal{C}$
%         \State $ \mathbf{q}^{M}_{c,n+1}\leftarrow Mask(\mathbf{q}_{c,n})$ \Comment{Cosine scheduled masking}
%       \EndFor
% \State $\mathbf{\tilde{q}}_c \leftarrow \mathbf{q}_{c,N}$
% \State \textbf{return} $\mathbf{\tilde{q}}_c$
%   \end{algorithmic}
% \end{algorithm}

\section{Experiments}
In this section, we evaluate the performance of our proposed model in prompt-based audio generation. To assess how well the model captures speaker consistency, the experiments were carried out in two scenarios: (1) the prompt speaker and the target speaker are the same, and (2) the prompt speaker and the target speaker are different. The second scenario is identical to the zero-shot voice conversion. Furthermore, we compare the runtime of our proposed model with the baseline. Our synthesized audio samples are publicly available at our demo page:~\url{https://jmhxxi.github.io/SoundGroup-demo/}.

\subsection{Experimental setup}
\subsubsection{Implementation details}
Our proposed model was trained for 800k iterations on 4 NVIDIA RTX8000 GPUs. The batch size was 128, with a gradient accumulation of 2. In this study, we use the open-sourced neural audio codecs from the AcademiCodec~\footnote{AcademiCodec:  \url{https://github.com/yangdongchao/AcademiCodec.}} toolkit. Acoustic tokenization was performed using HiFi-Codec, producing 50 frames per second, resulting in the target bitrate of $50\cdot4\cdot\log_{2}{1024}=2000$~bps. Our proposed model was trained with all of the training datasets of Libri-TTS~\cite{zen19_interspeech}, and the HiFi-Codec was pretrained with the same datasets as in the original configuration. For evaluation, we used the Libri-TTS test-clean subset so that all speakers in the evaluation set were unseen during training. We randomly selected the evaluation set consisting of 720 sentences and 20 sentences per speaker. All the speech data were sampled with a sampling rate of 24~kHz. For semantic tokenization, we used the pre-trained Wav2Vec~2.0 XLSR~\cite{babu2021xls}, with a total of 512 clusters for $k$-means clustering, and they were temporarily aligned to corresponding acoustic tokens. We evaluated the runtime on the single NVIDIA RTX8000 GPU to compare inference speed.

\subsubsection{Baselines}
We employed the SoundStorm model with the SoundStream codec as a baseline architecture. To eliminate data dependencies, we trained the SoundStream codec using the same dataset as HiFi-Codec. Following the SoundStorm configuration, we utilized the $6000$~bps SoundStream codec with $N_q=12$. We compared the SoundStorm and the proposed model by varying iteration numbers. During decoding, we used (16, 1, 1, ..., 1) iterations of SoundStorm for $N=27$ and greedy sampling for $N=12$.  Additionally, for different speaker prompt settings, we employed the variational inference-based (VITS)~\cite{kim2021conditional} voice conversion model as a baseline. We added ECAPA-TDNN~\cite{desplanques20_interspeech} as a reference encoder to the VITS for speaker conditioning. 

\subsubsection{Evaluation metrics}
We performed a Mean Opinion Score (MOS) test to assess synthesized speech quality, with 17 evaluators rating naturalness. To measure intelligibility, we computed the Character Error Rate~(CER) using a pretrained Whisper~\cite{radford2023robust} large model in official implementation. For speaker similarity, Similarity Mean Opinion Score~(SMOS) and Speaker Embedding Cosine Similarity~(SECS) were used. For SMOS evaluation, 17 listeners assessed how well the generated speech captured the speaker identity of the prompt speech. For SECS, we quantified the cosine distance between the speaker embeddings of the generated and prompt speech, using WavLM-TDNN~\cite{chen2022wavlm} as a pretrained speaker verification model.

\subsection{Results and Analysis}
% \begin{table}[th]
% \setlength{\tabcolsep}{2pt}
% \setlength{\arrayrulewidth}{0.2mm}
% \caption{Comparison of results for audio generation.}
% \label{tab:multi}
% \centering
% \begin{tabular}{l r r r r}
% \toprule
% \textbf{Method} & \multicolumn{1}{l}{\textbf{CER}}& \multicolumn{1}{l}{\textbf{SECS}}& \multicolumn{1}{l}{\textbf{\quad MOS}}& \multicolumn{1}{l}{\textbf{\; SMOS}} \\ 
% \midrule
% Ground Truth        & 4.1 & 0.632 & 4.72$\pm$0.06 & 4.65$\pm$0.07  \\
% \midrule
% SoundStorm (N=27)   & 6.1 & 0.212 &4.3$\pm$0.10&3.47$\pm$0.09  \\
% \midrule
% Proposed (N=27)  & 6.1 & 0.212 &4.3$\pm$0.10&3.47$\pm$0.09  \\
% Proposed (N=17)  & \textbf{6.1} & \textbf{0.212} &4.2$\pm$0.10 & \textbf{3.47$\pm$0.09}  \\
% Proposed (N=7)  & \textbf{6.1} & \textbf{0.212} &4.0$\pm$0.10 & \textbf{3.47$\pm$0.09}  \\
% \bottomrule
% \end{tabular}
% \end{table}

\begin{table}[th]

\setlength{\tabcolsep}{2pt}
\setlength{\arrayrulewidth}{0.2mm}

\caption{Comparison of results for audio generation. MOS and SMOS are described with 95\% confidence intervals.}
\centering
\begin{tabular}{l r r r r}
\toprule
\textbf{Method} & \multicolumn{1}{l}{\textbf{CER}}& \multicolumn{1}{l}{\textbf{SECS}}& \multicolumn{1}{l}{\textbf{\quad MOS}}& \multicolumn{1}{l}{\textbf{\; SMOS}} \\ 
\midrule

 \multicolumn{5}{c}{\textit{With Same Prompt Speaker}}\\
Ground Truth        & 0.71 & 0.737 & 4.55$\pm$0.08 & 4.55$\pm$0.07  \\
SoundStorm \textit{($N=12$)}   & 3.02 & 0.429 &3.24$\pm$0.07&3.99$\pm$0.07  \\
SoundStorm \textit{($N=27$)}   & 2.87 & 0.437 &3.83$\pm$0.07&4.10$\pm$0.08 \\
\midrule
Proposed \textit{($N_c=1,N=2$)}  & 2.90 & 0.412 &3.06$\pm$0.07 & 3.63$\pm$0.08 \\
Proposed \textit{($N_c=5,N=6$)}  & 2.69 & 0.475 &4.06$\pm$0.07 & 4.30$\pm$0.07  \\
Proposed \textit{($N_c=11,N=12$)}  & 2.57 & 0.476 &4.27$\pm$0.07 & 4.42$\pm$0.07  \\
Proposed \textit{($N_c=26,N=27$)}  & \textbf{2.36} & \textbf{0.487} &\textbf{4.49$\pm$0.06}& \textbf{4.47$\pm$0.06}  \\
\midrule
 \multicolumn{5}{c}{\textit{With Different Prompt Speaker}}\\

 VITS + Ref. \textit{(conversion)}& 2.71 & 0.382 &3.39$\pm$0.08&3.82$\pm$0.07\\
  SoundStorm \textit{($N=12$)} & 3.23 & 0.381 &3.54$\pm$0.08&3.86$\pm$0.08\\
  SoundStorm \textit{($N=27$)}  & 3.11 & 0.392 &3.86$\pm$0.08&4.24$\pm$0.07\\
  \midrule
 Proposed \textit{($N_c=1,N=2$)}  & 3.28 & 0.367 &3.10$\pm$0.07 & 3.58$\pm$0.08\\
 Proposed \textit{($N_c=5,N=6$)} & 2.53 & 0.406&4.11$\pm$0.07&4.22$\pm$0.07\\
 Proposed \textit{($N_c=11,N=12$)} & 2.49 & 0.416 &4.38$\pm$0.07&4.51$\pm$0.06\\
 Proposed \textit{($N_c=26,N=27$)} & \textbf{2.46} & \textbf{0.417} &\textbf{4.41$\pm$0.07}&\textbf{4.54$\pm$0.05}\\
 \bottomrule
 \label{audiogen}
 \vspace{-0.2cm}
\end{tabular}

\end{table}

\subsubsection{Prompt-based audio generation}
We present the results of prompt-based audio generation in Table~\ref{audiogen} where $N_c$ and $N$ indicate the number of iterations for coarse acoustic tokens and the total number of iterations, respectively. Our proposed model demonstrates superior performance in all metrics compared to the SoundStorm baseline at the same number of iterations, $N$. Moreover, our proposed model exhibits significantly better performance when compared to the VITS-based model in different prompt speaker setting. This indicates the proposed model's excellence in speaker similarity, audio quality, and speech intelligibility. Notably, in the case of SoundStorm, remarkable performance degradation was observed when $N$ was reduced to 12. In contrast, our proposed model maintained performance even with $N=6$, surpassing the performance of SoundStorm \textit{(N=27)}. The performance drop of the proposed model~\textit{(N=2)} was induced by the absence of G-IPD sampling, which failed to account for group-wise conditional dependency.
\subsubsection{Inference speed}
\begin{figure}
\centerline{\includegraphics[width=0.95\columnwidth]{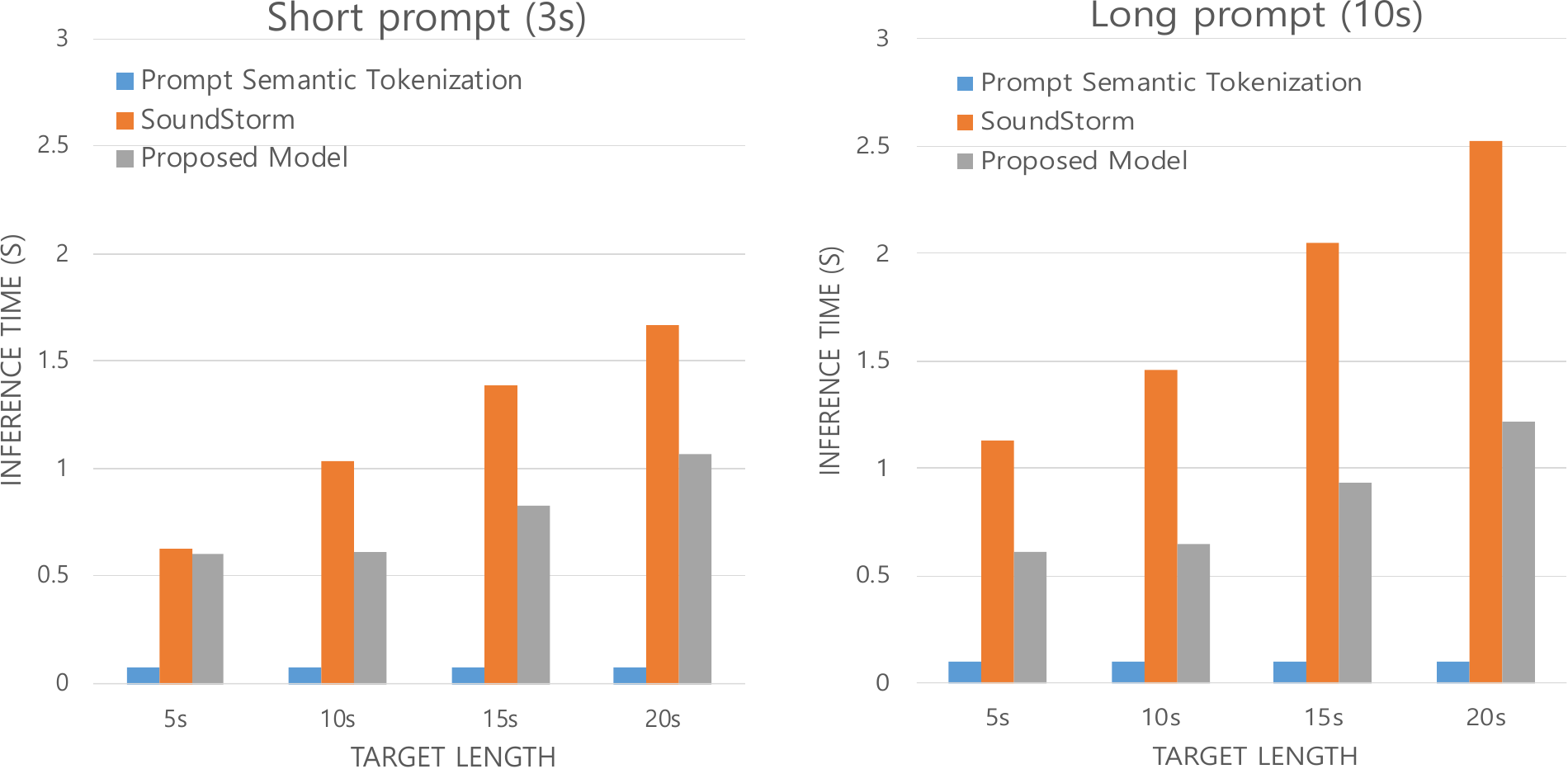}}
% \centerline{\includegraphics[width=0.80\columnwidth]{inference_speed.pdf}}
\caption{Comparison of inference speed. The prompt semantic tokenization is only used in SoundStorm's sampling process, and presented SoundStorm's runtime is evaluated without prompt semantic tokenization}
\label{inference_speed}
\vspace{-0.3cm}
\end{figure}

We compared the inference speed of our proposed model to that of SoundStorm. For a fair comparison, we fixed the total iteration number $N=27$. As shown in Fig.~\ref{inference_speed}, the proposed model is much faster than SoundStorm across all target and prompt lengths. Although the runtime of the self-attention module was dependent on the sequence length, our cross-attention-based architecture was less affected by the variations in prompt and target length. In particular, the runtime gap between proposed model and SoundStorm increase in long prompt setting, because proposed model avoids repetitive computation of prompt part throughout the inference process. As indicated in Table~\ref{audiogen}, we expect that our proposed model can reduce $N$ without a significant performance drop, resulting in much faster inference.

\section{Conclusion}

We have proposed a fast and high-quality codec language model for parallel audio generation using Group-Masked Language Modeling. For future work, we plan to extend our proposed model to support the zero-shot multi-speaker text-to-speech via a text-to-semantic translation model.

% \section*{Acknowledgment}

\bibliographystyle{IEEEtran}

\bibliography{mybib}

\end{document}